\documentclass[12pt, epsfig]{article}
\usepackage{graphicx}
  \usepackage{amsthm,amsfonts}
  \usepackage{amsmath}
  \usepackage{color}
  \usepackage{ulem}
\newcommand{\bea}   {\begin{eqnarray}}
\newcommand{\eea}   {\end{eqnarray}}
\def\hw{\widetilde{w}}

\def\hf{\frac{1}{2}}

\begin{document}
\renewcommand{\thefootnote}{\fnsymbol{footnote}}

\thispagestyle{empty}

\title{${\ell}$-oscillators from second-order invariant PDEs of the \\centrally extended Conformal Galilei Algebras}

\author{N. Aizawa\thanks{{E-mail: \textit{aizawa@mi.s.osakafu-u.ac.jp}}},\quad 
Z. Kuznetsova\thanks{{E-mail: \textit{zhanna.kuznetsova@ufabc.edu.br}}}
\quad and\quad F.
Toppan\thanks{{E-mail: \textit{toppan@cbpf.br}}}
\\
\\
}
\maketitle

\centerline{$^{\ast}$
{\it Department of Mathematics and Information Sciences,}}
{\centerline {\it\quad
Graduate School of Science, Osaka Prefecture University, Nakamozu Campus,}}
{\centerline{\it\quad Sakai, Osaka 599-8531 Japan.}}
\centerline{$^{\dag}$
{\it UFABC, Av. dos Estados 5001, Bangu,}}{\centerline {\it\quad
cep 09210-580, Santo Andr\'e (SP), Brazil.}
\centerline{$^{\ddag}$
{\it CBPF, Rua Dr. Xavier Sigaud 150, Urca,}}{\centerline {\it\quad
cep 22290-180, Rio de Janeiro (RJ), Brazil.}
~\\
\maketitle
\begin{abstract}

We construct, for any given ${\ell}=\frac{1}{2}+{\mathbb{N}}_0$, the second-order, linear PDEs which are invariant under the centrally extended Conformal Galilei Algebra. \par At the given ${\ell}$, two invariant equations in one time and ${\ell}+\frac{1}{2}$ space coordinates are obtained.
The first equation possesses a continuum spectrum and generalizes the free Schr\"odinger equation (recovered for ${\ell}=\frac{1}{2}$) in $1+1$ dimension. The second equation (the ``$\ell$-oscillator") possesses a discrete, positive spectrum. It generalizes the $1+1$-dimensional harmonic oscillator (recovered for $\ell=\frac{1}{2}$). The spectrum of the ${\ell}$-oscillator, derived from a specific $osp(1|2\ell+1)$ h.w.r., is explicitly presented.\par
The two sets of invariant PDEs are determined by imposing (representation-dependent) {\it on-shell invariant conditions} both for {\it degree} $1$ operators (those with continuum spectrum) and for {\it degree } $0$ operators
(those with discrete spectrum).\par
The on-shell condition is better understood by enlarging the Conformal Galilei Algebras with the addition of certain second-order differential operators. Two compatible structures (the algebra/superalgebra duality) are defined for the enlarged set of operators.

~\\\end{abstract}
\vfill

\rightline{%CBPF-NF-xxx/14
}

\newpage
\section{Introduction}

In this paper four results are presented. At first we enlarge the one-dimensional, centrally extended, Conformal Galilei Algebras (CGAs) of the half-integer, ${\ell}=\frac{1}{2}+{\mathbb N}_0$, series. The $sl(2)$ subalgebra elements and the central charge have an integer grading 
\textcolor{black}{with respect to the Cartan element of $ sl(2).$ } 
The remaining generators have a half-integer grading.  
The enlargement consists in adding to the set of generators the anticommutators of the generators with half-integer grading.  Two compatible structures are defined on the enlarged set of generators \cite{Toppan}, namely a (finite, non semi-simple) Lie algebra of commutators and a ${\mathbb Z}_2$-graded Lie
algebra of (anti)commutators. This compatibility is referred to as {\it algebra}/{\it superalgebra duality}.\par
The enlarged algebra/superalgebra possesses a differential realization induced by the differential realization
of the original CGA 
\textcolor{black}{
 (the differential realizations of the latter have been computed, e.g., in \cite{NdORM1,AiKiSe}). 
} 
The differential realization 
\textcolor{black}{suitable to the present work} 
requires one time and $\ell+\frac{1}{2}$ space coordinates. The first algebra of the series, obtained at $\ell=\frac{1}{2}$, is the Schr\"odinger algebra; since it is realized by just one space coordinate, this whole CGA series is referred to as ``one-dimensional".\par
The second result consists in imposing, for the given differential realization,  separate {\it on-shell invariant conditions} for the operators of grading $\pm 1$ ({\it degree } $\pm1$ operators) and grading $0$ ({\it degree} $0$ operators) of the enlarged algebra. By construction, these operators are second-order differential
operators. The operators which solve the {\it on-shell} condition induce an invariant PDE for the centrally extended CGA. At degree $1$ (or $-1$) we recover the invariant PDEs obtained in \cite{AiKiSe}. The novel feature is the invariant operator at degree $0$. The degree $1$ invariant equation, in one time and ${\ell }+\frac{1}{2}$ space coordinates, generalizes the free Schr\"odinger equation (recovered for ${\ell}=\frac{1}{2}$) in $1+1$ dimension.  Its associated ``static" equation possesses a continuum spectrum.  For the degree $0$ operator a transformation (which generalizes the one described in \cite{Nie73} and \cite{BluSht} for the Schr\"odinger case)
of the space and time coordinates (our third result) allows to recast the associated invariant PDE as a second-order differential equation in the new variables. In the new form it generalizes the $1+1$-dimensional harmonic oscillator (recovered for $\ell=\frac{1}{2}$).  The associated ``static" equation is given by a Hamiltonian which possesses a positive, discrete spectrum of eigenvalues. The last and main result of our paper is the presentation of this invariant ``${\ell}$-oscillator" Hamiltonian and the computation of its spectrum.
The latter  can be derived from a given highest weight (or lowest weight, depending on the conventions) representation of $osp(1|2\ell+1)$, which plays the role of the  spectrum-generating, {\it off-shell} symmetry subalgebra of the invariant PDE.\par
The constructions that we employed are explained at length in the text. \\ For convenience we present here the main results. The $\ell$-oscillator invariant equation is
\bea\label{losc1}
i\partial_\tau\Psi(\tau, {\bf x}_j) &=& {\bf H}^{(\ell)} \Psi(\tau, {\bf x}_j),
\eea
for the Hamiltonian ${\bf H}^{(\ell )}$ which, in a canonical form and suitably normalized, is given by
\bea\label{losc2}
{\bf H}^{(\ell)}&=&-\frac{1}{2m}\partial_{{\bf x}_1}^2+ \frac{m}{2}{{\bf x}_1}^2+\sum_{j=1}^{\ell-\frac{1}{2}}\left( (2j+1){\bf x}_{j+1}\partial_{{\bf x}_{j+1}} -(2\ell-2j+1){\bf x}_j\partial_{{\bf x}_{j+1}}\right)\nonumber\\
&&+\frac{1}{8}(2\ell -1)(2\ell+3). 
\eea
$\tau$ is the time coordinate (of scaling dimension $[\tau]=-1$), while the ${\bf x}_j$'s, $j=1,\ldots, {\ell}+\frac{1}{2}$,  are (anisotropic) space coordinates of scaling dimension
$[{\bf x}_j]= -j+\frac{1}{2}$.
\par
The spectrum of the ${\bf H}^{(\ell )}$ Hamiltonian, given by the energy eigenvalues $E_{\vec n}$, is
\bea\label{lspectrum}
E_{\vec n}&=&\sum_{j=1}^{\ell+\frac{1}{2}}\omega_j n_j + \omega_0,\quad{\text{with}}\quad \omega_j= (2j-1), \quad\omega_0=\frac{1}{8}(2\ell +1)^2.
\eea
In the above formula the $n_j$'s are non-negative integers, $\omega_j$ is the energy of the $j$-th oscillatorial mode and $\omega_0$ is the vacuum energy.
The spectrum coincides with the spectrum of ${\ell}+\frac{1}{2}$ decoupled oscillators of appropriate frequency.\par
The Hamiltonian is a second-order differential operator with respect to ${\bf x}_1$ and a first-order differential operator with respect to the remaining space coordinates. One should note that the space coordinates are coupled. The Hamiltonian of the Harmonic oscillator is recovered not only at ${\ell}=\frac{1}{2}$, but also from the consistent set of restrictions $\partial_{{\bf x}_j}\Psi(\tau,\vec{\bf x})=0$ for $j=2,\ldots, {\ell}+\frac{1}{2}$. \par It should be noted that ${\bf H}^{(\ell )}$, despite having a real,
positive spectrum, is not, for $\ell \geq \frac{3}{2}$, a Hermitian operator (a large class of non-Hermitian operators with real spectrum, $PT$-symmetric operators \cite{Bend1,Bend2}, pseudo-hermitian operators \cite{Pseudo1}, are currently very actively investigated).\par
The challenging problem of finding dynamical systems and invariant PDEs for the Conformal Galilei Algebras is much studied. In a series of papers \cite{GomiKami,GalaMas-PU,AndGaGoMa,Andr1,Andr2} it was shown that the Pais-Uhlenbeck oscillators are invariant under the ${\ell}=\frac{1}{2}+{\mathbb N}_0$ CGAs. These systems, however, unlike the ${\ell}$-oscillator (\ref{losc1}),  are defined by higher-derivatives equations. The ${\ell}$-oscillator induces a second-order differential equation.\par
The scheme of the paper is as follows. In Section {\bf 2} we introduce the one-dimensional centrally extended CGAs and present the ${\ell} =\frac{3}{2}$ differential realization. In Section {\bf 3} we introduce the
enlarged algebras and discuss the algebra/superalgebra duality. In Section {\bf 4} we introduce the on-shell condition to derive the invariant PDEs.  For ${\ell}=\frac{3}{2}$ we derive the explicit invariant operators at degree  $1$ and $0$. The transformation relating their suitable differential realizations, applied to the ${\ell}=\frac{3}{2}$ case, is presented
in Section  {\bf 5}. In Section {\bf 6} we explicitly discuss the spectrum-generating subalgebra for the ${\ell}=\frac{3}{2}$-invariant oscillator. Section {\bf 7} summarizes the previous constructions for the most
general, ${\ell}=\frac{1}{2}+{\mathbb N}_0$, case. For convenience, a table with the first ${\ell}$-oscillator Hamiltonians up to ${\ell }=\frac{9}{2}$ is given. In the Conclusions we present some open questions and the lines of future investigations.

\section{The centrally extended, $\ell\in\frac{1}{2}+\mathbb{N}_0$, CGAs}

We introduce here the Conformal Galilei Algebras (CGAs) in $d=1$ space dimensions \cite{NdORM1}.  
In the following we denote them as ``${\mathfrak{cga}}_{\ell}$". They are labelled by a parameter ${\ell}$ which is either a non-negative integer or a positive half-integer number.\par
The non-centrally extended  CGAs consist of an $sl(2)$ subalgebra with generators $z_{\pm1}, z_0$ ($z_0$ being the Cartan generator), acting on an abelian subalgebra ($2\ell+1$ generators $w_j$, $j=-{\ell}, -{\ell}+1,\ldots, {\ell}$, which span a spin-${\ell}$ representation of $sl(2)$). The generator $z_0$ induces a grading, so that $[z_{\pm 1}]=\pm 1, [z_0]=0, [w_j]=j$. For $\ell$ half-integer ($ {\ell}\in \frac{1}{2}+\mathbb{N}_0$), the corresponding CGA admits a central extension \cite{MarTach} (the centrally extended algebra will be denoted as ``${\widehat{\mathfrak{cga}}}_\ell$").  In the presence of the central extension $c$, the abelian subalgebra is replaced by $\ell+\frac{1}{2}$ pairs of Heisenberg subalgebras given by  $w_{\pm j}$. 
The ${\ell}=\frac{1}{2}$ case corresponds to the one-dimensional Schr\"odinger algebra, spanned by $6$ generators, including the central charge $c$. For reasons discussed in Section {\bf 3} we focus here on the centrally extended CGAs; therefore $\ell$ is a half-integer number. \par
The ${\widehat{\mathfrak{cga}}}_\ell$ algebra contains two generators of grading $0$ ($z_0$ and $c$), $\ell +\frac{3}{2}$ generators in the positive sector ($z_{+1}$ and $w_j$, with $j>0$) and an equal number of generators in the negative sector ($z_{-1}$ and $w_{-j}$, with $j>0$).\par
With standard techniques (see e.g. \cite{Helgason,Knapp,Dobrev}) one can construct a $D$-module rep realized by first-order differential operators induced by a coset construction. The differential operators depend on the space and time coordinates $t, x_j$, which are dual to the positive generators. The time $t$ is dual to $z_{+1}$ and has dimension $[t]=-1$, while the $x_j$'s are dual to the $w_{j}$'s. Therefore, $[x_j]= -j$. One should note the anisotropy of the space coordinates, as well as the fact that, for $\ell>\frac{1}{2}$ and in presence of the central extension, more than two coordinates 
%%%(the ``time" and the ``space") 
are required to realize the Conformal Galilei Algebra ${\widehat{\mathfrak{cga}}}_\ell$.\par
The $\ell=\frac{1}{2}$ Schr\"odinger algebra has been directly constructed as the symmetry algebra of the free Schr\"odinger equation in $1+1$-dimensions (see e.g. \cite{Nie}); its explicit presentation can be found in that work and several other papers in the literature. The next interesting and much less studied case corresponds to $\ell=\frac{3}{2}$. The algebra ${\widehat{\mathfrak{cga}}}_{\ell=\frac{3}{2}}$ is spanned by the $8$ generators $z_0, z_{\pm 1}, w_{\pm\frac{1}{2}}, w_{\pm {\frac{3}{2}}}, c$. Explicitly, its non-vanishing commutators are given by
\bea\label{commutators}
\relax[z_{+1},z_{-1}]&=& 2z_0,\nonumber\\
\relax [z_0, z_{\pm 1}] &=& \pm z_{\pm 1},\nonumber\\ 
\relax [z_0, w_{\pm\frac{3}{2}}]&=& \pm\frac{3}{2} w_{\pm \frac{3}{2}},\nonumber\\
\relax [z_0, w_{\pm\frac{1}{2}}]&=& \pm\frac{1}{2} w_{\pm \frac{1}{2}},\nonumber\\
\relax[z_{\pm 1}, w_{\pm \frac{1}{2}}]&=& w_{\pm \frac{3}{2}},\nonumber\\
\relax[z_{\pm 1}, w_{\mp \frac{1}{2}}]&=& 2 w_{\pm \frac{1}{2}},\nonumber\\
\relax[z_{\pm 1}, w_{\mp \frac{3}{2}}]&=& 3 w_{\mp \frac{1}{2}},\nonumber\\
\relax[w_{+\frac{1}{2}},w_{-\frac{1}{2}}] &=& c,\nonumber\\
\relax[w_{+\frac{3}{2}},w_{-\frac{3}{2}}] &=& \textcolor{black}{-}3c.
\eea
Its $D$-module representation, constructed, as explained above, in \cite{AiKiSe}, is realized by differential operators depending on $t,x_{\frac{1}{2}}, x_{\frac{3}{2}}$. For simplicity we will denote $x_{\frac{1}{2}}\equiv x$ and $x_{\frac{3}{2}}\equiv y$.
Their respective scaling dimensions are $[t]=-1, [x]=-\frac{1}{2}, [y]=-\frac{3}{2}$. \par
We overline the generators in the $D$-module rep. They are given by
\bea\label{threehalfdmod}
{\overline z}_{+1}&=& \partial_t,\nonumber\\
{\overline z}_{0}&=& -t\partial_t -\frac{1}{2}x\partial_x-\frac{3}{2}y\partial_y-1,\nonumber\\
{\overline z}_{-1}&=& -t^2\partial_t -tx\partial_x-3ty\partial_y-3y\partial_x-cx^2-2t,\nonumber\\
{\overline w}_{+\frac{3}{2}}&=& \partial_y,\nonumber\\
{\overline w}_{+\frac{1}{2}}&=& t\partial_y +\partial_x,\nonumber\\
{\overline w}_{-\frac{1}{2}}&=& t^2\partial_y+2t\partial_x+cx,\nonumber\\
{\overline w}_{-\frac{3}{2}}&=& t^3\partial_y+3t^2\partial_x+3tcx-3cy,\nonumber\\
{\overline c}&=& c.
\eea
This construction naturally extends to the ${\widehat{\mathfrak{cga}}}_\ell$ algebras for $\ell>\frac{3}{2}$. 

\section{Enlarged algebras: algebra/superalgebra duality}

For the Conformal Galilei Algebras in the half-integer series ($\ell=\frac{1}{2}+\mathbb{N}_0$), $z_0$ induces an integer grading on the generators $z_{\pm 1}, z_0, c$ and a half-integer grading on the $w_j$'s generators ($\forall g\in {\widehat{\mathfrak{cga}}}_\ell$, $[z_0, g] =n_g g$). It is therefore quite natural to interpret the grading induced  by $z_0$ as discriminating the even sector ${\cal G}_0$ (generators $g$ s.t. $n_g\in \mathbb{Z}$) from the odd sector
${\cal G}_1$ (generators $g$ s.t. $n_g\in \frac{1}{2}+\mathbb{Z}$) of an associated superalgebra.\par
By recalling that the $w_j$'s generators induce $\ell+\frac{1}{2}$ pairs of Heisenberg subalgebras, this possibility is made concrete by the existence (see \cite{Dic}) of an oscillatorial representation of the
$B(0,n)=osp(1|2n)$ superalgebra in terms of $n$ bosonic generators. Here $n=\ell+\frac{1}{2}$.\par
For consistency a closed superalgebra structure requires that the anticommutators of the odd generators belong to the even sector ${\cal G}_0$. Therefore we need to add to ${\cal G}_0$ the $(2\ell+1)(\ell+1)$
generators $w_{i,j} =\{w_i,w_j\}$ of grading $i+j$ (we have $w_{i,j}=w_{j,i}$). 
\par
The enlarged superalgebra spanned by $z_0, z_{\pm 1}, c, w_{i,j}\in {\cal G}_0$ and $w_{j}\in {\cal G}_1$ is a non semisimple, finite, closed, Lie superalgebra endowed with graded commutators and satisfying the graded Jacobi identities. It will be denoted as ``${\mathfrak s}{\widehat{\mathfrak{cga}}}_\ell$". It contains 
$2\ell^2+3\ell+5$ even generators and $2\ell+1$ odd generators. The generators $w_{i,j}$ close the
$sp(2\ell+1)$ bosonic subalgebra, while the set of $w_{i,j}, w_j$ generators close the $osp(1|2\ell+1)$ subalgebra. We can therefore write
\bea
{\mathfrak s}{\widehat{\mathfrak{cga}}}_\ell&=& u(1)\oplus sl(2)\oplus_S osp(1|2\ell +1),
\eea
where ``$\oplus_S$" denotes the semidirect sum.\\
In the superalgebra framework the central charge $c$ entering ${\widehat{\mathfrak{cga}}}_\ell$ is an extra $u(1)$ generator. 
%%%
It is worth pointing out that the construction of
the superalgebra $ {\mathfrak s}{\widehat{\mathfrak{cga}}}_\ell $ is based on a different viewpoint and requires a different procedure from the construction of the supersymmetric extensions of the CGAs which have been discussed in \cite{dh}---\cite{AizKuznTop}. Indeed, these supersymmetric extensions can be identified as symmetries of supersymmetric models, while  $ {\mathfrak s}{\widehat{\mathfrak{cga}}}_\ell $ is the symmetry superalgebra of a purely bosonic system.
%\cite{dh,SY1,SY2,AzLu,Sakaguchi,BaMa,FeLu,Mandal,mas,aiz,AizKuznTop}

The $D$-module representation of ${\widehat{\mathfrak{cga}}}_\ell$ in terms of first-order differential operators (as given in (\ref{threehalfdmod}) for $\ell =\frac{3}{2}$) is naturally extended to a realization of ${\mathfrak s}{\widehat{\mathfrak{cga}}}_\ell$ in terms of differential operators, with the operators ${\overline w}_{i,j}=\{{\overline w}_i,{\overline w}_j\}$ being of second order.\par
Once constructed the anticommutators $w_{i,j}$ from the set of $w_j$'s generators in ${\widehat{\mathfrak{cga}}}_\ell$ we can pose the question whether the ${\widehat{\mathfrak{cga}}}_\ell$ algebra, enlarged with the addition of the $w_{i,j}$ generators, closes as a non-semisimple finite Lie algebra (the brackets being defined by the ordinary commutators). The answer is positive. We denote as ``${\mathfrak{e}}{\widehat{\mathfrak{cga}}}_\ell$" the bosonic Lie algebra spanned by $z_{\pm 1}, z_0, c, w_j, w_{i,j}$. 
The algebra ${\mathfrak{e}}{\widehat{\mathfrak{cga}}}_\ell$ contains $2\ell^2+5\ell+6$ generators. They are all even.
We have, explicitly,
\bea
{\mathfrak{e}}{\widehat{\mathfrak{cga}}}_\ell &=& {\widehat{\mathfrak{cga}}}_\ell\oplus_S sp(2\ell+1).
\eea
Both ${\mathfrak{s}}{\widehat{\mathfrak{cga}}}_\ell$ and ${\mathfrak{e}}{\widehat{\mathfrak{cga}}}_\ell$
are obtained from ${\widehat{\mathfrak{cga}}}_\ell$ by adding an extra set of generators, expressed as anticommutators. In terms of the differential realizations given by (\ref{threehalfdmod}) and its $\ell>\frac{3}{2}$ counterparts, the second order differential operators ${\overline w }_{i,j}$ entering both ${\mathfrak{s}}{\widehat{\mathfrak{cga}}}_\ell$ and ${\mathfrak{e}}{\widehat{\mathfrak{cga}}}_\ell$ are identical.
Stated otherwise, on the same set of first and second-order differential operators, two mutually compatible structures can be defined. The first structure is the ${\mathbb Z}_2$-graded superalgebra ${\mathfrak{s}}{\widehat{\mathfrak{cga}}}_\ell$. The second structure is the ordinary Lie algebra ${\mathfrak{e}}{\widehat{\mathfrak{cga}}}_\ell$. We refer to this property of Conformal Galilei Algebras with half-integer $\ell$ as the algebra/superalgebra duality. It can be schematically expressed by the correspondence
\bea
{\mathfrak{s}}{\widehat{\mathfrak{cga}}}_\ell&\Leftrightarrow& {\mathfrak{e}}{\widehat{\mathfrak{cga}}}_\ell.
\eea
The symbol ``$\Leftrightarrow$" denotes this duality relation.

\section{Invariant PDEs from the on-shell condition} 

We address here the problem of constructing Partial Differential Equations (PDEs) admitting the centrally extended Conformal Galilei Algebras as their symmetry algebras. One should note that this is an inverse problem with respect to the derivation of the Schr\"odinger algebra (${\widehat{{\mathfrak{cga}}}}_\ell$, for $\ell=\frac{1}{2}$) from the (free) Schr\"odinger equation.
In that case the invariant equation is assumed and its symmetry algebra, induced by first-order differential operators, is derived with standard techniques (see \cite{Ovsia,Olver}).  For ${\ell }\geq \frac{3}{2}$ a reverse problem has to be solved. The first-order differential operators generating ${\mathfrak{cga}}_\ell$ are known. We have instead to determine the invariant PDEs associated with the given $D$-module reps. \par
%%%
One possibility is offered by the method, based on Lie symmetries, which produce non-linear invariant PDEs,
see \cite{FusCher,CherHen}. 
Another possibility, leading to linear invariant PDEs, is based on the construction of singular vectors of the 
given representation \cite{Dobrev} (see \cite{AiKiSe} for the case of ${\widehat{{\mathfrak{cga}}}}_\ell$). 
We discuss in this Section a different approach to determine linear invariant PDEs, based on imposing an
on-shell invariant condition (see \cite{FN94} and \cite{Nie}).  
Applied to the ${\widehat{{\mathfrak{cga}}}}_\ell$ algebras, this approach naturally leads to second-order linear PDEs (invariant PDEs with higher derivatives can, \textcolor{black}{in principle}}, also be constructed).\par
Let us consider, for a given ${\ell}$, the enlarged algebras introduced in Section {\bf 3} (either the superalgebra
${\mathfrak{s}}{\widehat{\mathfrak{cga}}}_\ell$ or its dual bosonic counterpart 
${\mathfrak{e}}{\widehat{\mathfrak{cga}}}_\ell$).  We can consider the most general generators $\Omega_r$ of grading $r$ (also called the \textit{degree} and defined by the commutator $[z_0,\Omega_r]=r\Omega_r$),
with $r=0, \pm 1$.  The $\Omega_r$'s are even generators entering both ${\mathfrak{s}}{\widehat{\mathfrak{cga}}}_\ell$ and $ {\mathfrak{e}}{\widehat{\mathfrak{cga}}}_\ell$. For ${\ell}=\frac{3}{2}$
we have, e.g.,
\bea\label{cohomology}
\Omega_{\pm 1} &=& a_1 z_{\pm 1}+a_2 w_{\frac{\pm 1}{2},\frac{\pm 1}{2}}+a_3w_{\frac{\pm 3}{2},\frac{\mp 1}{2}},\quad \quad a_{1,2,3}\in \mathbb{C},\nonumber\\
\Omega_0 &=& b_0 c+b_1 z_{0}+b_2 w_{\frac{1}{2},\frac{-1}{2}}+b_3w_{\frac{3}{2},\frac{-3}{2}},\quad b_{0,1,2,3}\in \mathbb{C}.
\eea
By construction, $\forall g\in {\mathfrak{s}}{\widehat{\mathfrak{cga}}}_\ell\Leftrightarrow {\mathfrak{e}}{\widehat{\mathfrak{cga}}}_\ell$, the commutators 
\bea
[g,\Omega_r]&=& \omega_r^g
\eea
close on the elements $\omega_r^g\in {\mathfrak{s}}{\widehat{\mathfrak{cga}}}_\ell\Leftrightarrow {\mathfrak{e}}{\widehat{\mathfrak{cga}}}_\ell$. \par

In a given differential realization of  ${\mathfrak{s}}{\widehat{\mathfrak{cga}}}_\ell\Leftrightarrow {\mathfrak{e}}{\widehat{\mathfrak{cga}}}_\ell$ the generators $\Omega_r$ are expressed as ${\overline\Omega}_r$,
while the r.h.s. elements $\omega_r^g$ are expressed as ${\overline \omega}_r^g$. \par
We have now all the ingredients to define a cohomological problem. To determine whether it is possible to choose ${\overline \Omega}_r$ in the $r$-graded sector of ${\mathfrak{s}}{\widehat{\mathfrak{cga}}}_\ell\Leftrightarrow {\mathfrak{e}}{\widehat{\mathfrak{cga}}}_\ell$ such that,
$\forall g$, the differential operators ${\overline \omega}_r^g$ are expressed as
\bea\label{cohom}
{\overline \omega}_r^g &=& f^g{\overline \Omega}_r,
\eea
where, for a given $g$, $f^g$ is a specific function of the space and time coordinates (for $\ell=\frac{3}{2}$, $f^g\equiv f^g(t,x,y)$) and $f^g$ can also be vanishing.
\par
As shown below, this cohomological problem admits non-trivial solutions in the presence of a non-vanishing central extension $c$.\par
The requirement that the Equation (\ref{cohom}) should be satisfied for any $g$ will be called the \textit{on-shell condition} for an invariant PDE. Indeed, if (\ref{cohom}) is satisfied for a given $r$, we obtain, applied to the solutions $\Psi(t,\vec{x})$ of
the partial differential equation
\bea
{\overline \Omega}_r\Psi(t,\vec{x})&=&0,
\eea 
the symmetries of the equation of motion
\bea
[{\overline g}, {\overline \Omega}_r]\Psi(t,\vec{x})&=&0.
\eea
We can express this property through the position 
\bea
[{\overline g}, {\overline \Omega}_r]=f^g{\overline \Omega}_r &\Rightarrow& 
[{\overline g}, {\overline \Omega}_r]\approx 0.
\eea
The existence of the on-shell symmetry implies that explicit solutions (eigenfunctions and eigenvalues) of the invariant PDE can be constructed in terms of the associated spectrum generating subalgebras (either belonging to ${\mathfrak{e}}{\widehat{\mathfrak{cga}}}_\ell$ or to ${\mathfrak{s}}{\widehat{\mathfrak{cga}}}_\ell$).\par
Let us present now the construction, from the on-shell condition, of the invariant PDE at $r=1$ induced by the 
differential realization (\ref{threehalfdmod}) of the ${\ell}=\frac{3}{2}$ case.
\par
For ${\ell }=3/2$ we have 
\bea
{\mathfrak s}{\widehat{\mathfrak{cga}}}_{\frac{3}{2}}= u(1)\oplus sl(2)\oplus_S osp(1|4)\quad&,&\quad
{\mathfrak{e}}{\widehat{\mathfrak{cga}}}_\ell = {\widehat{\mathfrak{cga}}}_{\frac{3}{2}}\oplus_S sp(4).
\eea
A non-trivial solution of the (\ref{cohom}) on-shell condition is guaranteed if we take the following linear combinations of generators
\bea\label{omegas}
\Omega_0 &=& z_0 +\frac{1}{4c}w_{\frac{1}{2},\frac{-1}{2}}-\frac{1}{4c}w_{\frac{3}{2},\frac{-3}{2}},\nonumber\\
\Omega_1&=& z_{+1} +\frac{1}{2c}w_{\frac{3}{2},\frac{-1}{2}}-\frac{1}{2c} w_{\frac{1}{2},\frac{1}{2}}.
\eea
The $D$-module rep (\ref{threehalfdmod}) of ${\widehat{\mathfrak{cga}}}_{\frac{3}{2}}$ induces
the second order differential operators ${\overline \Omega}_0$ and ${\overline\Omega}_1$, explicitly given by
\bea{\overline{ \Omega}}_0= -t {\overline{\Omega}}_1\quad&,&\quad
{\overline{\Omega}}_{1}= \partial_t+x\partial_y-{\frac{1}{c}} {\partial_x}^2.
\eea
In the given differential realization ${\overline \Omega}_1$ satisfies the
on-shell condition (\ref{cohom}). Indeed, 
$\forall g\in {\mathfrak{s}}{\widehat{\mathfrak{cga}}}_{\frac{3}{2}}\Leftrightarrow {\mathfrak{e}}{\widehat{\mathfrak{cga}}}_{\frac{3}{2}}$, its only non-vanishing commutators are given by
\bea\label{onshell-omega1}
\relax [{\overline z}_{-1}, {\overline\Omega}_1] &=&-2{\overline \Omega}_0 ~= 2t{\overline\Omega}_1,\nonumber\\
\relax [{\overline z}_0, {\overline \Omega}_1]&=& {\overline \Omega}_1.
\eea
Therefore $f^{z_{-1}}=2t$, $f^{z_0}=1$, $f^g=0$ otherwise.\par
For what concerns the ${\overline \Omega}_0$ operator, its commutators are vanishing when applied to
the solutions of the
\bea\label{freethreehalfeom}
{\overline\Omega}_1\Psi(t,x,y)=0&\equiv&\Psi_t(t,x,y) +x\Psi_y(t,x,y) -\frac{1}{\textcolor{black}{c}}\Psi_{xx}(t,x,y)=0
\eea
invariant PDE. \par
We have, indeed, that its only non-vanishing commutators in the basis expressed by
${\overline z}_{\pm 1}, {\overline z}_0, {\overline w}_j, {\overline w}_{i,j}, {\overline c}$ are
given by
{\textcolor{black}{\bea\label{omega0}
\relax [{\overline z}_{+1}, {\overline \Omega}_0]&=& -{\overline \Omega}_{1} \quad=t^{-1}{\overline\Omega}_0,\nonumber\\
\relax [{\overline z}_{-1},{\overline \Omega}_0]&=& -t^2{\overline \Omega}_1~ = t{\overline\Omega}_0.
\eea
}}}
We have, furthermore,
\bea\label{Omega0-1}
\relax [{\overline \Omega}_1, {\overline \Omega}_0]&=& -{\overline \Omega}_1.
\eea
It follows, from the last equations in the r.h.s. of (\ref{omega0}), that ${\overline{\Omega}}_0$ satisfies the on-shell condition for a singular choice of space-time functions,
namely
$f^{z_{+1}}=t^{-1}$, $f^{z_{-1}}=t$ and $f^g=0$ otherwise.\par
The (\ref{freethreehalfeom}) invariant PDE is the counterpart, at ${\ell }=\frac{3}{2}$, of the free Schr\"odinger equation in $1+1$ dimensions.  {\textcolor{black}{The centralizer algebra for the operator ${\overline{\Omega}}_1$ (induced by the operators $\overline g$ which strictly satisfy $[{\overline g}, {\overline \Omega}_1]=0$), will be called the  \textit{off-shell} invariant algebra for the (\ref{freethreehalfeom}) PDE. This subalgebra is obtained by disregarding the $sl(2)$ Borel generators ${\overline z}_0, {\overline z_{-1}}$.} In the superalgebra framework the off-shell invariant algebra can be presented as
\bea\label{offshellfree}
u(1)\oplus u(1)\oplus_S osp(1|4) &\subset & {\widehat{\mathfrak{scga}}}_{\frac{3}{2}},
\eea
with the $u(1)$ generator acting on $osp(1|4)$ given by ${\overline z}_{+1}$. \par
The PDE (\ref{freethreehalfeom}), derived from the on-shell condition, 
is identical to the lowest member of the hierarchy, whose construction is based on singular vectors, given in \cite{AiKiSe}. The on-shell condition allowed us to identify here another invariant operator (${\overline\Omega}_0$, at degree $0$), whose importance is discussed in the next Section.

\section{The invariant PDE of degree $0$}
\label{Sec:oscil}

Besides the free case, in $1+1$ dimensions the Schr\"odinger algebra is derived as the symmetry algebra
of the Schr\"odinger equation for two other choices of the potential, the linear potential and the quadratic potential of the harmonic oscillator \cite{Nie73,Nie74,Boyer}. The difference between the free case and the oscillator case
lies in the fact that the time-derivative operator is, in the first case, associated with a positive root of the $sl(2)$ subalgebra, as in formula (\ref{threehalfdmod}). In the second case it is associated with the Cartan generator (for the linear potential, the time-derivative operator is a symmetry generator which does not coincide with a generator of the $sl(2)$ subalgebra \cite{Boyer, Toppan}). In the framework of the on-shell condition, the free Schr\"odinger equation is the invariant PDE at degree $1$, while the equation of the harmonic oscillator is the invariant PDE at degree $0$ \cite{Toppan}. The appearance of a discrete spectrum for the harmonic oscillator in contrast to the continuous spectrum of the free particle can be traced to these differences. \par
At ${\ell}=\frac{1}{2}$ the $D$-module rep associated with the harmonic oscillator can be recovered from the original $D$-module rep
via a transformation (this point has been discussed in \cite{Nie73} and \cite{BluSht}). 
This transformation can be extended to other values of ${\ell}$ entering
${\widehat{\mathfrak{gca}}}_{\ell}$. We present it for ${\ell}=\frac{3}{2}$. Essentially, the transformation requires presenting $z_0$ as the time-derivative operator with respect to a new time variable (in a related context, see \cite{LinTop}). \par
We recall that we denoted as ${\overline g}$, a $g\in{\widehat{\mathfrak{gca}}}_{\frac{3}{2}}$ generator in the 
$D$-module rep (\ref{threehalfdmod}). We denote as ``${\widetilde g}$" the generator in the new $D$-module rep.
%which, with abuse of language, can be referred to as the \textit{oscillatorial $D$-module rep}. 
We can write ${\overline g}\in {\overline V}$, ${\widetilde g}\in {\widetilde V}$, where ${\overline V}, {\widetilde V}$ are the corresponding $D$-module reps. The transformation $\tau$ mapping
\bea
\tau: {\overline V}\rightarrow {\widetilde V}\quad &,& \quad \tau: {\overline g} \mapsto {\widetilde g},
\eea
can be realized in three steps:\par
\textit{i})  at first any given generator ${\overline g}$ in (\ref{threehalfdmod}) is dressed by the similarity transformation ${\overline g} \mapsto {\check g} = t{\overline g}t^{-1}$;\par
\textit{ii}) next, the ${\check g}$ generators are reexpressed as differential operators in the new variables
$s,u,v$, related to the previous variables $t,x,y$ through the positions
\bea
t&=& e^s,\nonumber\\
x&=& e^{\frac{s}{2}}u,\nonumber\\
y&=& e^{\frac{3s}{2}} v,
\eea
($s$ plays now the role of the new ``time" variable);\par
\textit{iii}) finally, the ${\check g}$ generators are dressed by a similarity transformation
which preserves ${\check z}_0$ as the time-derivative operator. There is an arbitrariness in the choice of the similarity transformation. In this and the next Section it is convenient to work with the choice
${\check g}\mapsto {\widetilde g} = e^{\frac{c u^2}{2}} \check{g}e^{-\frac{c u^2}{2}}$. A different similarity transformation, leading to the ``canonical" form of the ${\ell}$-oscillator Hamiltonian given by (\ref{losc2}), is introduced in Section {\bf 7}.
\par
The result of the three combined operations produces a $D$-module rep of ${\widehat{\mathfrak{cga}}}_{\frac{3}{2}}$, given by the first-order differential operators in $s,u,v$,
\bea\label{oscdmodrep}
{\widetilde z}_{+1}&=& e^{-s}(\partial_s-\frac{1}{2}u\partial_u-\frac{3}{2}v\partial_v-1 +\frac{1}{2}cu^2),\nonumber\\
{\widetilde  z}_{0}&=& -\partial_s,\nonumber\\
{\widetilde z}_{-1}&=&e^s( -\partial_s-\frac{1}{2}u\partial_u-\frac{3}{2}v\partial_v -3v\partial_u-\frac{1}{2}cu^2-1+3cuv),\nonumber\\
{\widetilde w}_{+\frac{3}{2}}&=& e^{-\frac{3 s}{2}}\partial_v,\nonumber\\
{\widetilde w}_{+\frac{1}{2}}&=&e^{-\frac{s}{2}}(\partial_v+ \partial_u-cu),\nonumber\\
{\widetilde w}_{-\frac{1}{2}}&=&e^{\frac{s}{2}}(\partial_v+2\partial_u-cu) ,\nonumber\\
{\widetilde w}_{-\frac{3}{2}}&=& e^{\frac{3s}{2}}(\partial_v+ 3\partial_u -3cv).\nonumber\\
{\widetilde c}&=& c.
\eea
In this differential realization, the generators $\Omega_0$, $\Omega_1$ introduced in (\ref{omegas})
are expressed as
\bea\label{oscinvop}
{\widetilde {\Omega}}_0 &=& -\partial_s-u\partial_v-\frac{3}{2}u\partial_u+\frac{3}{2}v\partial_v+
\frac{1}{\textcolor{black}{c}}{\partial_u}^2+\frac{1}{2} cu^2,
\nonumber\\
{\widetilde{ \Omega}}_1 &=& -e^{-s}{\widehat \Omega}_0. \label{Omega-Osci}
\eea
Both ${\widetilde\Omega}_0$ (at degree $r=0$) and ${\widetilde\Omega}_1$ (at degree $r=1$) satisfy the on-shell condition (\ref{cohom}). Indeed, their respective non-vanishing commutators with the operators in ${\widetilde V}$ are given by  
\bea \label{onshell-tilder1}
\relax [{\widetilde z}_{+1}, {\widetilde \Omega}_0] &=& e^{-s}{\widetilde\Omega}_0,\nonumber\\
\relax [{\widetilde z}_{-1}, {\widetilde \Omega}_0] &=& e^s{\widetilde\Omega}_0
\eea
and
\bea \label{onshell-tilder2}
\relax [{\widetilde z}_{0}, {\widetilde \Omega}_1] &=& {\widetilde \Omega}_1,\nonumber\\
\relax [{\widetilde z}_{-1}, {\widetilde \Omega}_1] &=& 2e^s{\widetilde \Omega}_1.
\eea
We have, furthermore, the relation
\bea\label{onshell-tilder3}
\relax[{\widetilde\Omega}_0,{\widetilde \Omega}_1]&=&{\widetilde\Omega}_1.
\eea
One should note that the degree $0$ on-shell invariant operator ${\widetilde \Omega}_0$, in this $D$-module rep, does not present an explicit dependence on the time coordinate $s$. Its associated invariant PDE is given by
\bea\label{oscthreehalfeom}
{\widetilde \Omega}_0\Psi(s,u,v)=0\quad &\equiv& \quad -\Psi_s -u\Psi_v-\frac{3}{2}u\Psi_u+\frac{3}{2}v\Psi_v+\frac{1}{\textcolor{black}{c}}\Psi_{uu}+\frac{1}{2}cu^2\Psi=0.
\eea
It is a second-order partial differential equation, containing a term proportional to $u^2$, which implements  the ${\ell}=\frac{3}{2}$ counterpart of the harmonic oscillator in $1+1$ dimensions. Its off-shell invariant algebra is obtained by disregarding the root generators ${\widetilde z}_{\pm 1}$. In the superalgebra framework it can be expressed as a subalgebra
\bea\label{offshellosc}
u(1)\oplus u(1)\oplus_S osp(1|4) &\subset & {\widehat{\mathfrak{scga}}}_{\frac{3}{2}}.
\eea
It is a different ${\widehat{\mathfrak{scga}}}_{\frac{3}{2}}$ subalgebra with respect to (\ref{offshellfree}). In that case the $u(1)$ generator acting on $osp(1|4)$ is the root generator $z_{+1}$, while here it is the Cartan generator $z_0$.

\section{Eigenfunctions and eigenvalues from the spectrum generating subalgebras}
\label{Sec:Eigen}

Both the $\ell =\frac{3}{2}$ degree $1$ (\ref{freethreehalfeom}) and degree $0$ (\ref{oscthreehalfeom}) invariant PDEs admit the $osp(1|4)$ superalgebra as a subalgebra of their respective off-shell symmetry algebras.\par
We focus here in the degree $0$ case. As already recalled, the superalgebras of the $osp(1|2n)$ series present a realization in terms of $n$ bosonic oscillators. In our case we implemented concretely this realization with $w_{\frac{\pm 1}{2}}$, $w_{\frac{\pm 3}{2}}$ expressed as first-order differential operators, see (\ref{oscdmodrep}). The superalgebra $osp(1|4)$ can be used as the spectrum generating algebra to construct
eigenfunctions and eigenvalues of the (\ref{oscthreehalfeom}) PDE. Based on the algebra/superalgebra duality discussed in Section {\bf 3} and in analogy with the construction for the ordinary harmonic oscillator in ($1+1$) dimensions, see \cite{Toppan}, two compatible viewpoints can be adopted here. In the bosonic viewpoint the construction of eigenfunctions and eigenvalues is derived from the Fock space of two bosonic oscillators. In the superalgebra viewpoint, the same result is recovered from a highest weight representation of the $osp(1|4)$ superalgebra.\par
{\textcolor{black}{The operator ${\widetilde{\Omega}}_0$ given in (\ref{oscinvop}) commutes with ${\widetilde {z}}_0$.}}
{\textcolor{black}{If we set
\bea 
{{\widetilde \Omega}}_0 &=& {\widetilde {z}}_0 +{\widetilde H},\nonumber\\
{\widetilde H }&=& \frac{1}{2c}(\hw_{-\hf} \hw_{+\hf} - \hw_{-\frac{3}{2}} \hw_{+\frac{3}{2}} ) + 1
=-u\partial_v-\frac{3}{2}u\partial_u+\frac{3}{2}v\partial_v+
\frac{1}{\textcolor{black}{c}}{\partial_u}^2+\frac{1}{2} cu^2,
\eea
we have that $[{\widetilde z}_0, {\widetilde H}]=0$. Therefore the equation ${\widetilde{\Omega}}_0\Psi(s,u,v)=0$ is solved by the common eigenfunctions $\Psi_E(s,u,v)$ such that
\bea\label{spliteq}
{\widetilde z}_0\Psi_E(s,u,v)= - {E} \Psi_E(s,u,v)\quad&,&\quad{\widetilde H}\Psi_E(s,u,v)= {E} \Psi_E(s,u,v).
\eea
${\widetilde H}$ plays the role of an effective Hamiltonian and $E$ is the energy level. Eigenstates and eigenvalues are obtained from a highest weight representation by applying the creation operators ${\widetilde w}_{\frac{-3}{2}}, {\widetilde w}_{\frac{-1}{2}}$ on the vacuum solution
$\psi_{vac}(s,u,v)$, defined  by the conditions
\bea\label{vacuum1}
&{\widetilde w}_{\frac{1}{2}}\psi_{vac}(u,v)={\widetilde w}_{\frac{3}{2}}\psi_{vac}(u,v)=0&
\eea
and 
\bea\label{vacuum2}
&-{\widetilde z}_0\psi_{vac}(s,u,v)={\widetilde H}\psi_{vac}(s,u,v)=E_{vac} \psi_{vac}(s,u,v).&
\eea
Due to (\ref{oscdmodrep}), the unnormalized solution of (\ref{vacuum1}) is 
\begin{equation}
  \psi_{vac} =\chi(s) e^{\hf cu^2}, \label{GSfunction}
\end{equation}
for an arbitrary function $\chi(s)$. The vacuum energy is given by $E_{vac}= 1$.}}\par
{\textcolor{black}{The first equation in (\ref{vacuum2}) constraints $\chi(s) $ to be $\chi(s)\propto e^s$.}}\par
{\textcolor{black}{
One can easily verify that the solutions of (\ref{spliteq}) can be expressed as $\psi_E(s,u,v)\propto e^{Es}\varphi_E(u,v)$.
The ground state function $ \varphi_{vac}(u,v)= e^{\frac{1}{2}cu^2} $ turns out to be independent of $v$ and normalizable in $(-\infty,\infty) $ if the central charge is restricted to $ c < 0$. }}\par
The higher energy eigenstates 
\bea 
\psi_{m,n} (s,u,v)&=&e^{sE_{m,n}}\varphi_{m,n}(u,v)
\eea of (\ref{spliteq}) are given by
\bea
  \psi_{m,n} &=& (\hw_{-\frac{3}{2}})^m (\hw_{-\hf})^n \psi_{vac}
\eea
(we can therefore also set $\psi_{vac}\equiv\psi_{0,0}$, as well as $\varphi_{vac}\equiv \varphi_{0,0}$).\par
Due to (\ref{vacuum1}) and to the (\ref{commutators}) commutators, the associated $E_{m,n}$ eigenvalues are
\bea\label{spectrum}
E_{m,n} &=& \frac{3}{2}m+\frac{1}{2} n + 1.
\eea 
One should note that they do not depend on the value of the central charge. Furthermore, the eigenvalues $E_{m,n}$ are degenerate for $E_{m,n}\geq \frac{5}{2}$.\par
Since ${\widetilde H}$ does not depend on $s$, the two-variable functions $\varphi_{m,n}(u,v)$ are solutions of the ``static" equation 
\bea\label{static}
{\widetilde H}\varphi_{m,n} (u,v)&=&E_{m,n}\varphi_{m,n}(u,v).
\eea
Oscillatorial solutions are recovered if $s$ is assumed to be imaginary. In terms of the new time variable $\tau$ we have
\bea
\psi_{m,n} (\tau,u,v) &=& e^{-i\tau E_{m,n}}\varphi_{m,n}(u,v), \quad\quad (s=-i\tau).
\eea
{\textcolor{black}{The first few  (unnormalized) eigenstates corresponding to the lowest energy eigenvalues (for $E_{m,n}\leq 3$) are given by
\bea\label{eigenstatesthreehalf}
E=1:&&\psi_{(0,0)}= e^se^{\frac{1}{2}cu^2},\nonumber\\
E=\frac{3}{2}:&&\psi_{(0,1)}=e^{\frac{3}{2}s}(cu e^{\frac{1}{2}cu^2}),\nonumber\\
E=2:&&\psi_{(0,2)}= e^{2s}(c(2 \textcolor{black}{+}cu^2)e^{\frac{1}{2}cu^2}),\nonumber\\
E=\frac{5}{2}:&&\psi_{(1,0)}= e^{\frac{5}{2}s}(3c(u-v)e^{\frac{1}{2}cu^2}),\nonumber\\
E=\frac{5}{2}:&&\psi_{(0,3)}= e^{\frac{5}{2}s}( \textcolor{black}{c^2u(6+cu^2)} e^{\frac{1}{2}cu^2}),\nonumber\\
E=3:&&\psi_{(1,1)}= e^{3s}(3c(1+cu(u-v))e^{\frac{1}{2}cu^2}),\nonumber\\
E=3:&&\psi_{(0,4)}= e^{3s}( \textcolor{black}{ c^2(12+12cu^2+c^2u^4) }e^{\frac{1}{2}cu^2}). 
\eea
By construction, the two-variable functions $\varphi_{m,n}(u,v)$ which solve the static equation (\ref{static}) are expressed as products of polynomials in $u,v$ which multiply the  ground state function $\varphi_{0,0}=e^{\frac{1}{2}cu^2}$.  }}

\section{The ${\ell}$-oscillator for any ${\ell}=\frac{1}{2}+{\mathbb N}_0$}

Differential realizations of $ {\widehat{\mathfrak{cga}}}_\ell $ for any half-integer $\ell$ have been computed in \cite{AiKiSe}. In that paper the second-order differential operators which, in our language, are on-shell invariant operators of degree $1$, were presented. These operators possess a continuum spectrum.\par
On the other hand, as shown in the previous Sections, a discrete spectrum can be obtained if we impose the on-shell condition for
degree $0$ operators. We discussed at length the derivation of  the (\ref{spectrum}) discrete spectrum for ${\ell}=\frac{3}{2}$. \par
We present here the general case of second-order invariant operators with discrete spectrum (the $\ell$-oscillators) for any ${\ell}=\frac{1}{2}+{\mathbb N}_0$.  Since the derivation is a straightforward generalization of the $\ell=\frac{3}{2}$ case, we can keep the discussion short.\par
The differential realizations of $ {\widehat{\mathfrak{cga}}}_\ell $ given in \cite{AiKiSe} require
the differential operators to depend on $t$ and the $ \ell + \hf $ variables $ x_{\hf}, x_{\frac{3}{2}}, \dots, x_{\ell}. $  The scaling dimension of the variables are $ [t] = -1, \ [x_j] = -j. $ 
A convenient change of notation, $ x_j \to y_{j+\hf} $, is here introduced.  The new variables are $ (t, y_a), $ with $ a = 1, 2, \dots, \ell + \hf. $  
In these new variables the generators given in \cite{AiKiSe} read as follows:
\begin{eqnarray} \label{anyl-free-real}
  \overline{z}_{+1} &=& \partial_t, \nonumber \\
  \overline{z}_0 &=& -t \partial_t - \sum_{a=1}^{\ell+\hf} (a -\hf )  y_a \partial_{y_a} - \delta,
  \nonumber \\
  \overline{z}_{-1} &=& 2t \overline{z}_0 + t^2 \partial_t - \sum_{a=1}^{\ell-\hf} (\ell+a+\hf ) y_{a+1} \partial_{y_a} 
  - (\ell+\hf ) \frac{c}{2} y_1^2,
  \nonumber  \\
  \overline{w}_j &=& \sum_{k=0}^{\ell-j} \binom{\ell-j}{k} t^{\ell-j-k} \partial_{y_{\ell+\hf-k}},
  \nonumber  \\
  \overline{w}_{-j} &=& \sum_{k=0}^{\ell-\hf} \binom{\ell+j}{k} t^{\ell+j-k} \partial_{y_{\ell+\hf-k}}
  - \frac{ (\ell+j)! c} {( \ell-\hf )! ( \ell+\hf )!} 
  \sum_{a=1}^{j+\hf} (-1)^a \frac{ (\ell+\hf-a)! }{ (j+\hf-a)! } t^{j+\hf-a} y_a,
  \nonumber \\
{\overline c} &=& c,
\end{eqnarray}
where $ j = \hf, \frac{3}{2}, \dots, \ell $ and $ \delta=\frac{1}{4} ( \ell + \hf )^2. $ 

 The  on-shell invariant second order operator $ \overline{\Omega}_1 $ is given by
\begin{equation}
  \overline{\Omega}_1 = \partial_t + \sum_{a=1}^{\ell-\hf} (\ell+\hf-a ) y_a \partial_{y_{a+1}}
  - \frac{1}{2c} ( \ell + \hf ) \partial_{y_1}^2. 
  \label{OnSehllOp-anyl}
\end{equation}
The corresponding degree $0$ operator can be defined by 
$ \overline{\Omega}_0 = -t \overline{\Omega}_1. $ 
It is straightforward to check that these two operators 
satisfy the same on-shell relations as  (\ref{onshell-omega1}), (\ref{omega0}) and (\ref{Omega0-1}).  
\par
For any half-integer ${\ell}$, the new differential realization, allowing to conveniently express the ${\ell}$-oscillator Hamiltonian, is obtained by performing the following three-step transformation on the operators entering (\ref{anyl-free-real}) and its associated enlarged algebra:

\textit{i}) 
the similarity transformation $ \overline{g} \mapsto \check{g} = t^{\delta}\, \overline{g}\, t^{-\delta} $
is applied to any such operator  ${\overline g}$;

\textit{ii}) a change of variables, $ (t,y_a) \mapsto (s, u_a) $, is performed on the differential operators $ \check{g}$,
\begin{equation}
  t = e^s, \qquad y_a = e^{ (a - \hf) s} u_a;
\end{equation}

\textit{iii}) the similarity transformation $ \check{g} \mapsto \widetilde{g} = \exp( -\frac{\lambda}{2}u_1^2 ) \check{g} \exp( \frac{\lambda}{2}u_1^2 ) $,
with $ \displaystyle \lambda = -\frac{c}{2\ell+1}$, is applied (for this choice of $\lambda$ one eliminates in the final $\ell$-oscillator Hamiltonian a term proportional to $u_1\partial_{u_1}$).
  \par

 The combination of these three operations produces  the following differential realization of  $ {\widehat{\mathfrak{cga}}}_\ell $ for any half-integer $\ell$: 
\begin{eqnarray}\label{generators-anyl}
  \widetilde{z}_{+1} &=& e^{-s} \Big( \partial_s - \sum_{a=1}^{\ell+\hf} ( a - \hf ) u_a \partial_{u_a} 
  + \frac{c}{2} \frac{1}{2\ell+1} u_1^2 - \delta \Big),
  \nonumber \\
  \widetilde{z}_{0} &=& - \partial_s,
  \nonumber \\
  \widetilde{z}_{-1} &=& e^s \Big( - \partial_s - \sum_{a=1}^{\ell+\hf} ( a - \hf ) u_a \partial_{u_a} 
  - \sum_{a=1}^{\ell-\hf} (\ell+\hf+a ) u_{a+1} \partial_{u_a} 
  + \frac{c}{2} (  \frac{1}{2\ell+1} - \ell - \hf ) u_1^2
  \nonumber \\
  && +
    \frac{c}{2}\, \frac{2\ell+3}{2\ell+1} u_1 u_2 - \delta \Big),
  \nonumber \\
  \widetilde{w}_{j} &=& e^{-js} \sum_{k=0}^{\ell-j} \binom{\ell-j}{k} \partial_{u_{\ell+\hf-k}} 
   - \delta_{j,\hf} \frac{c}{2\ell+1} e^{-\frac{s}{2}}  u_1,
  \nonumber \\
  \widetilde{w}_{-j} &=& e^{js} \Big(
     \sum_{k=0}^{\ell-\hf} \binom{\ell+j}{k} \partial_{u_{\ell+\hf-k}} 
     - \frac{ (\ell+j)! c } { \big( \ell-\hf \big)! \big(\ell+\hf \big)! } 
     \sum_{a=1}^{j+\hf} (-1)^a \frac{(\ell+\hf-a)!}{ (j+\hf-a)! } u_a
  \nonumber \\
  && -
   \frac{c}{2\ell+1}  \binom{\ell+j}{\ell-\hf}  u_1
  \Big),\nonumber\\
{\widetilde c}&=& c.
\end{eqnarray}
The on-shell invariant operators $ \overline{\Omega}_0$, ${\overline{\Omega}}_1$  are mapped into
\begin{eqnarray} \label{tildeOmegas-anyl}
  \widetilde{\Omega}_0 &=& 
    -\partial_s + \sum_{j=2}^{\ell+\hf} ( j-\hf ) u_j \partial_{u_j} 
    - \sum_{j=1}^{\ell-\hf} (\ell+\hf-j ) u_{j} \partial_{u_{j+1}}
    \nonumber \\
    &&+ 
     \frac{1}{2c} ( \ell + \hf ) \partial_{u_1}^2 
     - \frac{c}{4} \frac{1}{2\ell+1} u_{1}^2 + \frac{1}{16}(2\ell-1) (2\ell+3),
   \nonumber \\
   \widetilde{\Omega}_1 &=& -e^{-s} \widetilde{\Omega}_0. 
\end{eqnarray}
The relations (\ref{onshell-tilder1}), (\ref{onshell-tilder2}) and (\ref{onshell-tilder3}) 
hold true for any half-integer $\ell.$  
\par
 We are able to reexpress, with the further change of variable $ s = -2i \tau $  and after setting $ c = -(2\ell+1) m$, 
the on-shell invariant PDE $ \widetilde{\Omega}_0 \psi(s,u_a) = 0 $ as
\begin{eqnarray}
 i \partial_{\tau} \psi(\tau,u_a) &=& {\bf H}^{(\ell)} \psi(\tau,u_a), 
 \nonumber \\
  {\bf H}^{(\ell)} &=& - \frac{1}{2m} \partial_{u_1}^2 + \frac{m}{2} u_1^2 + \sum_{a=2}^{\ell+\hf} (2a-1) u_a \partial_{u_a}
 - \sum_{a=1}^{\ell-\hf} (2\ell+1-2a) u_a \partial_{u_{a+1}} 
 \nonumber \\
 & &  + \frac{1}{8}(2\ell-1) (2\ell+3).
 \label{InvPDEforl}
\end{eqnarray}
It coincides (after setting  $u_a={\bf x}_a$ to improve readability) with the ${\ell}$-oscillator equation that we introduced in equations (\ref{losc1}) and (\ref{losc2}).\par
 Since the differential operator $ {\bf H}^{(\ell)} $ is independent of $ \tau $, the PDE (\ref{InvPDEforl}) 
admits solutions of the form
\begin{equation}
  \psi(\tau,u_a) = e^{-iE\tau} \varphi(u_a), \qquad {\bf H}^{(\ell)} \varphi(u_a) = E \varphi(u_a). 
\end{equation}
The eigenvalue problem for the operator $ {\bf H}^{(\ell)} $ is solved via an algebraic method. 
The vacuum solution $ \varphi_{vac}(u_a) $ is defined to satisfy
\begin{equation}
  \widetilde{w}_j \,\varphi_{vac}(u_a) = 0, \quad j = \hf, \frac{3}{2}, \dots, \ell .
\end{equation}
We get, as a consequence,
\begin{equation}
 \varphi_{vac}(u_a) = \exp ( -\frac{m}{2} u_1^2 ), \qquad 
 {\bf H}^{(\ell)} \varphi_{vac}(u_a) = 2\delta \varphi_{vac}(u_a). 
\end{equation}
The relations
\begin{equation}
  [\widetilde{\Omega}_0, \widetilde{w}_{\pm j} ] = 0, \qquad 
  [\widetilde{z}_0, \widetilde{w}_{\pm j}] = \pm j \widetilde{w}_{\pm j}, \qquad
  2\widetilde{\Omega}_0 = 2 \widetilde{z}_0 + {\bf H}^{(\ell)},
\end{equation}
imply that 
\begin{equation}
  [{\bf H}^{(\ell)}, \widetilde{w}_{\pm j} ] = \mp 2j \widetilde{w}_{\pm j}.
\end{equation}
Therefore, the eigenvalues and eigenfunctions are respectively given by
\begin{eqnarray}
   E_{\vec{n}} &=& \sum_{a=1}^{\ell+\hf} (2a-1) n_a + \hf ( \ell + \hf )^2,
   \nonumber \\
   \varphi_{\vec{n}} &=& \widetilde{w}_{-\hf}^{n_1} \widetilde{w}_{-\frac{3}{2}}^{n_2} \cdots \widetilde{w}_{-\ell}^{n_{\ell+\hf}} \varphi_{vac}(u_a),
\end{eqnarray}
where $ \vec{n} = (n_1, n_2, \dots, n_{\ell+\hf}) $ is a ($\ell+\hf$)-component vector with entries
$n_a\in{\mathbb N}_0$. \par
This concludes the derivation of  the discrete spectrum that we introduced in (\ref{lspectrum}).\par
The table of the first ${\ell}$-oscillator Hamiltonians for $\ell \leq \frac{9}{2}$ is here reported for convenience. With the same conventions as in the Introduction (namely $u_a={\bf x}_a$), we have
\bea
{\bf H}^{(\frac{1}{2})}&=& -\frac{1}{2m}\partial_{{\bf x}_1}^2+ \frac{m}{2}{{\bf x}_1}^2,\nonumber\\
{\bf H}^{(\frac{3}{2})}&=&-\frac{1}{2m}\partial_{{\bf x}_1}^2+ \frac{m}{2}{{\bf x}_1}^2+3{\bf x}_2\partial_{{\bf x}_2} -2{\bf x}_1\partial_{{\bf x}_2}+\frac{3}{2},\nonumber\\
{\bf H}^{(\frac{5}{2})}&=&-\frac{1}{2m}\partial_{{\bf x}_1}^2+ \frac{m}{2}{{\bf x}_1}^2+ 3{\bf x}_2\partial_{{\bf x}_2}+5{\bf x}_3\partial_{{\bf x}_3}  -4{\bf x}_1\partial_{{\bf x}_2}-2{\bf x}_2\partial_{{\bf x}_3}+4,\nonumber\\
{\bf H}^{(\frac{7}{2})}&=& -\frac{1}{2m}\partial_{{\bf x}_1}^2+ \frac{m}{2}{{\bf x}_1}^2+3{\bf x}_2\partial_{{\bf x}_2} +5{\bf x}_3\partial_{{\bf x}_3} +7{\bf x}_4\partial_{{\bf x}_4} -6{\bf x}_1\partial_{{\bf x}_2}-4{\bf x}_2\partial_{{\bf x}_3}-2{\bf x}_3\partial_{{\bf x}_4}+\frac{15}{2},\nonumber\\
{\bf H}^{(\frac{9}{2})}&=& -\frac{1}{2m}\partial_{{\bf x}_1}^2+ \frac{m}{2}{{\bf x}_1}^2+3{\bf x}_2\partial_{{\bf x}_2}+5{\bf x}_3\partial_{{\bf x}_3} +7{\bf x}_4\partial_{{\bf x}_4} +9{\bf x}_5\partial_{{\bf x}_5} \nonumber\\
&& -8{\bf x}_1\partial_{{\bf x}_2}-6{\bf x}_2\partial_{{\bf x}_3}-4{\bf x}_3\partial_{{\bf x}_4}-2{\bf x}_4\partial_{{\bf x}_5}+12.
\eea
\section{Conclusions}

For ${\ell}\geq\frac{3}{2}$, the ${\ell}$-oscillator Hamiltonians fit into the class of non-Hermitian operators with real spectrum (see, e.g., \cite{Bend1,Pseudo1}). 
Besides being real, their spectrum is positive and discrete. It coincides with the spectrum of a set of decoupled harmonic oscillators of appropriate ($\nu_j\propto 2j-1$) frequency. The full quantum theory of the ${\ell}$-oscillators requires the introduction of an inner product, norm, orthogonality conditions for the polynomials such as
those entering (\ref{eigenstatesthreehalf}), etc. It deserves being fully scrutinized in a separate paper.\par
The spectrum is recovered from the ${\ell}+\frac{1}{2}$ harmonic oscillators realizing the $osp(1|2\ell+1)$ off-shell invariant subalgebra. In the ${\ell}\rightarrow\infty$ limit an infinite tower of oscillatorial modes
is created. The connection of the $sp(2m)$ algebras and their orthosymplectic extensions with the higher spin theories (see, e.g., \cite{Fro, Vas}) is well-established. It is tempting, in the light of the AdS/CFT correspondence and non-relativistic holography, to conjecture that ${\ell}$-oscillators could appear in the dual, CFT side of higher-spin theories (possibly, in some non-relativistic contraction limit). This is an important point for future investigations.\par 
The Pais-Uhlenbeck oscillators (see, e.g.,  \cite{AndGaGoMa}) are invariant under the CGAs. Moreover, up to a normalization factor, their frequencies coincide with the ${\ell}$-oscillator spectrum of frequencies. The two systems on the other hand are quite different. The Pais-Uhlenbeck oscillators describe higher-derivatives theories, while the
${\ell}$-oscillators are second-order (no higher derivative) PDEs. A possible connection can arise from the on-shell condition. We investigated it for the differential realizations belonging to the enlarged ${\widehat{\mathfrak{cga}}}_{\ell}$ algebras. This is the natural setup for second-order invariant PDEs.
One can of course search for solutions of the on-shell condition for differential realizations of
${\cal U}({\widehat{\mathfrak{cga}}}_{\ell})$, the ${\widehat{\mathfrak{cga}}}_{\ell}$ Universal Enveloping Algebra. If such solutions are encountered, then the associated invariant PDEs are of higher order.\par
Our construction can be straightforwardly extended to more general cases. It can be applied to $d$-dimensional CGAs with
half-integer ${\ell}$. The first invariant PDEs of the series, at ${\ell}=\frac{1}{2}$, are either the free Schr\"odinger or the harmonic oscillator equation in $1+d$-dimensions. For generic ${\ell}=\frac{1}{2}+{\mathbb N}_0$ one recovers, as off-shell invariant subalgebras, both $so(d)$ and $osp(1|d(2\ell+1))$
 (realized by $d(\ell+\frac{1}{2})$ oscillators). The fact that the oscillatorial modes can be accommodated into spin representations of $so(d)$ makes quite interesting to investigate the possible arising of Regge trajectories for ${\ell}$-oscillators with $d>1$.\par
The last comment regards supersymmetry. The supersymmetric extensions of ${\widehat{\mathfrak{cga}}}_{\ell}$ possess a ${\mathbb Z}_2$-grading. Their enlarged superalgebras require a second ${\mathbb Z}_2$-grading induced by the Cartan element of the $sl(2)$ subalgebra. The {\it algebra}/{\it superalgebra} duality of the enlarged bosonic algebra is replaced, in the supersymmetric case, by the {\it superalgebra}/${\mathbb Z}_2\times {\mathbb Z}_2$-{\it graded algebra} duality of the enlarged superalgebra. The ${\mathbb Z}_2\times {\mathbb Z}_2$-graded algebras have been investigated, albeit to a less extent than superalgebras,  in the mathematical literature, see e.g. \cite{RittWyl,Scheu}. A work on the
{\it superalgebra}/${\mathbb Z}_2\times {\mathbb Z}_2$-{\it graded algebra} duality is currently under finalization.
\\ {~}~
\par {\Large{\bf Acknowledgments}}
{}~\par{}
%~\par
N. Aizawa is grateful to UFABC, where this research was conducted, for invitation and warm hospitality.
This research was supported by FAPESP under Grant 2014/03560-0 and by CNPq under PQ Grant 306333/2013-9.

\end{document}